\def\lae{\mathrel{<\kern-1.0em\lower0.9ex\hbox{$\sim$}}}
\def\gae{\mathrel{>\kern-1.0em\lower0.9ex\hbox{$\sim$}}}

\def\z{\ensuremath{z_{850}} }
\def\g{\ensuremath{g_{475}} }

\documentclass{emulateapj}

\slugcomment{Accepted for Publication in ApJS}
\shortauthors{{JORD\'AN ET AL.}}
\shorttitle{ACSFCS I}

\begin{document}

\title{The ACS Fornax Cluster Survey. I. 
Introduction to the Survey and Data Reduction Procedures\altaffilmark{1}}

\author{Andr\'es Jord\'an\altaffilmark{2}, 
John P. Blakeslee\altaffilmark{3},
Patrick C\^ot\'e\altaffilmark{4}, 
Laura Ferrarese\altaffilmark{4},
Leopoldo Infante\altaffilmark{5},
Simona Mei\altaffilmark{6},
David Merritt\altaffilmark{7},
Eric W. Peng\altaffilmark{4},
John L. Tonry\altaffilmark{8},
Michael J. West\altaffilmark{9,10}}

\begin{abstract}

The Fornax Cluster is a conspicuous cluster
of galaxies in the southern hemisphere and the second largest collection
of early-type galaxies within $\lae 20$ Mpc after the Virgo
Cluster. 
In this paper,
we present a brief introduction to the ACS Fornax Cluster
Survey --- a program to image, in the F475W ($g_{475}$)
and F850LP ($z_{850}$) bandpasses, 43 early-type galaxies
in Fornax using the {\it Advanced Camera
for Surveys} (ACS) on the {\it Hubble Space Telescope}.
Combined with a companion survey of Virgo, the ACS Virgo Cluster Survey,
this represents the most comprehensive imaging survey to date 
of early-type galaxies in cluster environments
in terms of depth, spatial resolution, sample size and homogeneity.
We describe the selection of the program galaxies, their basic
properties, and the main science objectives of the survey
which include the measurement of luminosities, colors
and structural parameters for globular clusters 
associated with these galaxies, an analysis of their
isophotal properties and surface brightness profiles,
and an accurate calibration of the surface
brightness fluctuation distance indicator. Finally, we 
discuss the data reduction procedures adopted for the survey.
\end{abstract}

\keywords{galaxies: clusters: individual (Fornax)
--- galaxies: elliptical and lenticular, cD
--- galaxies: distances and redshifts
--- galaxies: star clusters
--- galaxies: nuclei
--- methods: data analysis
}

\altaffiltext{1}{Based on observations with the NASA/ESA 
{\it Hubble Space Telescope}
obtained at the Space Telescope Science Institute, which is operated 
by the Association
of Universities for Research in Astronomy, Inc., under 
NASA contract NAS 5-26555}
\altaffiltext{2}{European Southern Observatory, 
Karl-Schwarzschild-Str. 2, 85748 Garching bei M\"unchen, Germany; ajordan@eso.org}
\altaffiltext{3}{Department of Physics and Astronomy,
Washington State University, 1245 Webster Hall, Pullman, WA 99163-2814}
\altaffiltext{4}{Herzberg Institute of Astrophysics, Victoria, 
BC V9E 2E7, Canada}
\altaffiltext{5}{Departamento de Astronom\'{\i}a y Astrof\'{\i}sica, 
Pontificia Universidad Cat\'olica de Chile, 
Avenida Vicu\~na Mackenna 4860, Casilla 306, Santiago 22, Chile}
\altaffiltext{6}{GEPI, Observatoire de Paris, Section de Meudon, 5 place Jules Janssen, 
92195 Meudon Cedex, France}
\altaffiltext{7}{Department of Physics, Rochester Institute of Technology,
84 Lomb Memorial Drive, Rochester, NY 14623}
\altaffiltext{8}{Institute for Astronomy, University of Hawai'i, 
2680 Woodlawn Drive, Honolulu, HI 96822}
\altaffiltext{9}{Department of Physics \& Astronomy, University of Hawai'i, 
Hilo, HI 96720}
\altaffiltext{10}{Gemini Observatory, Casilla 603, La Serena, Chile}

\section{Introduction}
\label{sec:intro}

Much of our current understanding of galaxy formation and
evolution is based on observations of galaxies in 
cluster environments.
While not necessarily representative of clusters at high redshift, 
nearby clusters have played a key role in shaping 
this understanding for the simple reason that they are amenable to study in
a level of detail that will never be possible for the more distant systems.

Among the nearby clusters, Virgo is almost certainly
the most thoroughly studied. Indeed, it can probably be said that
Virgo is the most closely examined cluster of galaxies
in the entire universe (see, e.g., the discussions in Huchra
1985; Binggeli, Sandage \& Tammann 1985; Binggeli 1999; Gavazzi et~al. 2003;
C\^ot\'e et~al. 2004).
After Virgo, the largest concentration of galaxies 
within $\la 20$ Mpc of our galaxy is the Fornax cluster.
This cluster, the most conspicuous
concentration of galaxies in the southern hemisphere, is centered
roughly at $\alpha \sim 3^{\mathrm h}35^{\mathrm m}$ and 
$\delta \sim -35.7\degr$ (Ferguson 1989a). 
It is located in the Fornax constellation, which was introduced in the 18th
century by the Abb\'e Nicolas Louis de la Caille while
mapping the southern skies from the Cape of Good Hope 
during the years 
1751--1753\footnote{The Fornax constellation was originally named ``Le Fourneau'', later
latiniced to ``Fornax Chimiae'' 
(Chemical Furnace). In the spirit of the illustration
de la Caille named several constellations after instruments of the
liberal and philosophical arts. He gave us also 'Apparatus
Sculptoris' (sculptor's apparatus, now Sculptor) 
and 'Antlia Pneumatica' (air pump, now Antlia), among others. See 
Evans (1951) for a brief account of La Caille's work in the Cape.}.

To the best of our knowledge, it was first identified
as a bonafide galaxy cluster by Shapley (1943) who compiled data on many
of the most prominent members and wrote
``Table 3 lists some of the information we have at hand concerning 
 a score of bright objects in the constellation 
Fornax, which are so located with respect to one another that the law of chance
is hard pressed if these objects are only accidentally near together. They appear
to constitute a real colony of galaxies, mutually operating.''. 
However, a complete census of cluster members was still many years away, and
even as late as 1956, \object[NGC 1399]{NGC~1399} --- the luminous cD/E0 galaxy at the
dynamical center of the main component of the cluster 
(Drinkwater, Gregg \& Colless 2001a) --- was listed
as a non-cluster galaxy in the catalog of Humason, Mayall \& Sandage (1956).

Studies of the Fornax cluster have often followed on the heels
of similar studies of the Virgo cluster. 
For instance, one of the first studies
of Fornax reported the discovery of dwarf galaxies
(Hodge 1959) after a similar population had been
 uncovered in Virgo by Reaves (1956).
The earliest catalogs of Fornax cluster members (Hodge 1960, Hodge, Pyper
\& Webb 1965) were eventually supplanted by the Fornax
Cluster Catalog (FCC) of Ferguson (1989a), who surveyed 40 square degrees using large-scale 
plates from the 2.5m Las Campanas telescope --- the same telescope that was used
some years before to assemble the Virgo Cluster Catalog (Binggeli et~al. 1985). The FCC remains the most comprehensive catalog of Fornax
currently available.

The globular cluster systems of the brightest Fornax galaxies were first tentatively detected around
\object[NGC 1374]{NGC~1374}, \object[NGC 1379]{NGC~1379}, 
and \object[NGC 1399]{NGC~1399} by carrying out starcounts in a single
deep photographic plate (Dawe \& Dickens 1975), a full two decades after a similar
population had been first discussed for \object[M87]{M87}, the giant elliptical in the center
of the Virgo cluster (Baum 1955). A systematic study of globular cluster systems
in Fornax was later presented
by Hanes \& Harris (1986) as one of a series of papers
on extragalactic globular clusters (which had included the study 
of globular clusters in Virgo galaxies in previous installments).

In the all-important task of establishing the cosmological distance ladder, Cepheids
were used by the HST Key Project on the Extragalactic Distance Scale to derive a 
distance to Fornax (Madore et~al. 1999) shortly after
the same program had identified Cepheids in Virgo 
(Freedman et~al. 1994; Ferrarese et~al. 1996).
Following initial feasibility tests of the surface brightness fluctuations 
(SBF; Tonry \& Schneider 1988) distance indicator on \object[M32]{M32} and 
\object[NGC 3379]{NGC~3379}, the method was applied to a significant number of galaxies in the 
Virgo Cluster (Tonry, Ajhar \& Luppino 1989, 1990). However, because of complications
arising from the significant line-of-sight depth of Virgo, the first accurate calibration
of the SBF method had to await subsequent observations of galaxies in the 
Fornax cluster (Tonry 1991).

Despite their shared proximity to our Galaxy, the two clusters show some obvious differences
that invite intercomparisons (see Table~\ref{tab:comp}, which summarizes the
basic properties of the clusters). It is clear both Fornax and Virgo offer a unique opportunity to 
examine the possible effects of environment on the properties
of cluster galaxies. For instance, Fornax is far more regular in shape, and
probably more dynamically evolved, than its
northern counterpart. It is also
considerably smaller and denser than Virgo, with a core radius $\sim 40$\% 
that of Virgo and a central density twice as large. The total
mass of Fornax is $\approx 7 \times 10^{13} M_\odot$, as estimated from
galaxy radial velocities (Drinkwater et~al. 2001a), which is $\sim$ 1/10 that of Virgo.
Fornax is 
thus more representative of the groups and poor clusters in which most galaxies
in the universe reside. 

\begin{deluxetable*}{lccc}
%\tabletypesize{\scriptsize}
\tablecaption{Basic Data for the Virgo and Fornax Clusters.\label{tab:comp}}
\tablewidth{0pt}
\tablehead{
\colhead{Property} & 
\colhead{Virgo} & 
\colhead{Fornax} & 
\colhead{References}\\
\colhead{(1)} &
\colhead{(2)} &
\colhead{(3)} &
\colhead{(4)} 
}
\startdata
Richness Class & 1 & 0 & 1,2\\
B-M Type & III & I & 1,2\\
Mass        & (4--7)$\times 10^{14} M_\odot$ & $(7\pm2)\times10^{13} M_\odot$ & 3,4,5 \\
Distance (Mpc) & 16.5 & 19.3 & 6,7\\
$\langle v_r \rangle$ (km sec$^{-1}$) & $1094 \pm 42$& $1493\pm36$& 5,8\\
$\sigma_v$ (km sec$^{-1}$) & $760$& $374\pm26$& 5,8\\
$r_c$ (Mpc) & $\approx$ 0.6 & $\approx$ 0.25 & 9\\
$n_0$ (gal Mpc$^{-3})$ & $\approx$ 250 & $\approx$ 500 & 9\\
$N$  & 1170 & 235& 9\\
$f_{\rm E+dE+S0+dS0}$ & 0.8 & 0.87& 9\\
$\langle kT \rangle_X$ (keV) & $2.58\pm0.03$ & $1.20\pm0.04$ &10\\
$\langle {\rm Fe} \rangle_X$ (solar) &  $0.34\pm0.02$& $0.23\pm0.03$ & 10\\
\enddata
\tablecomments{
Key to columns---(1) Cluster properties listed in the Table are the richness class, 
Bautz-Morgan (B-M) Type, mass, distance, average heliocentric radial velocity $\langle v_r \rangle$,
velocity dispersion $\sigma_v$, King model core radius $r_c$, central galaxy density $n_0$, 
number of members $N$ with $B \la 18$ and within $3.5r_c$,  the fraction of these
that are E, dE, S0 or dS0 ($f_{\rm E+dE+S0+dS0}$), and 
the average temperature, $\langle kT \rangle_X$, 
and Fe  abundance, $\langle {\rm Fe} \rangle_X$,  of the intracluster medium as derived
from X-ray observations (excluding the inner cluster regions);
(2-3) Value of given property for the Virgo and Fornax clusters respectively;
(4) Reference for quoted value:
1.- Abell, Corwin \& Olowin 1989; 
2.- Girardi et~al. 1995; 
3.- McLaughlin 1999; 
4.- Tonry et~al. 2000; 
5.- Drinkwater et~al. 2001b; 
6.- Mei et~al. 2007; 
7.- Tonry et~al. 2001; 
8.- Binggeli, Sandage \& Tamman 1987; 
9.- Ferguson 1989b; 
10.- Fukazawa et~al. 1997. 
}
\end{deluxetable*}

Being a much  more compact cluster than Virgo, Fornax is an ideal target for the
calibration of distance indicators (see above). However, like most clusters, it
does show some substructure --- the main
subcluster is centered on \object[NGC 1399]{NGC~1399}, 
while a subcluster that includes \object[NGC 1316]{NGC~1316} (= Fornax A)
is centered $3^\circ$ to the southwest (Drinkwater et~al. 2001a). Within the main
subcluster, there seems to be an infalling clump associated with 
\object[NGC 1404]{NGC~1404}, a picture
that is supported by the characteristics of its X-ray emission 
(Scharf et~al. 2005, Machacek et~al. 2005).

The installation of the Advanced Camera for Surveys (ACS; Ford et~al. 1998) on board
the {\it Hubble Space Telescope} (HST) improved dramatically the telescope's imaging 
capabilities. In the first cycle of operations with ACS, we initiated the ACS Virgo 
Cluster Survey (ACSVCS; C\^ot\'e et~al. 2004), a program to obtain deep F475W
($\approx$ Sloan $g$) and F850LP ($\approx$ Sloan $z$) ACS/WFC images for 100
early-type members of the Virgo cluster.
As the second step in a broader program to obtain high-resolution imaging for a large,
well-defined sample
of early-type galaxies in nearby clusters, we initiated the ACS Fornax Cluster Survey (ACSFCS).
This ACS/WFC survey of 43 early-type galaxies in the Fornax cluster was carried out 
using the same observational strategy that was employed in the
ACSVCS. Collectively, these two surveys provide homogeneous $\g$-
and $\z$-band imaging of 143
galaxies with magnitudes $B_T \lae 16$, a dataset with
wide-ranging scientific potential.

In this paper, the first in a series, we present an introduction to the ACSFCS and discuss
the data reduction procedures adopted for it.
The paper is organized as follows: we present a brief description of the survey objectives
in \S\ref{sec:sciobj}, outline selection of sample galaxies in 
\S\ref{sec:selec} and describe the observations in \S\ref{sec:obs}.
We detail the data reduction procedures of the
survey in \S\ref{sec:data_red} and conclude in~\S\ref{sec:summary}.

\section{Motivation and Objectives}
\label{sec:sciobj}

The motivations for undertaking a high-spatial resolution, multi-wavelength
imaging survey of galaxies in nearby clusters has been presented
in the context of the ACSVCS by C\^ot\'e et~al. (2004). The survey images
have a wide-ranging use for a variety of scientific applications, some 
of which were foreseen during the planning of the survey and some which 
were not. At the time of writing, the broad array of scientific results from
the ACSVCS have been reported in sixteen refereed 
papers\footnote{See the ACSVCS website: 
http://www.cadc.hia.nrc.gc.ca/community/ACSVCS}.

As mentioned above, the Fornax cluster provides a unique counterpart
to Virgo in terms of environment; e.g., studies that were
carried out in Virgo can be extended to Fornax, both increasing the total sample size and allowing
the search for effects that cluster environment may have on findings based on the richer
and less dense Virgo cluster. 
It is not our intention to provide a thorough review of the background of each 
science objective of the present survey. Instead, 
we will briefly discuss four main scientific topics that will be addressed using this
new dataset. The interested reader is referred to the published papers from the ACSVCS for 
detailed references and results.

\subsection{Extragalactic Globular Clusters}

A key objective of the ACSFCS is the study of thousands
of globular clusters (GCs) associated with the program galaxies. The images 
are sufficiently deep that $\sim 90\%$ of the GCs falling within the ACSFCS fields
can be detected at a high level of completeness (C\^ot\'e et~al. 2004).
Moreover, the high 
spatial resolution of {\it HST} allows the measurement the half-light radii for GCs
by fitting point spread function (PSF)-convolved models to the two-dimensional light distributions 
(Jord\'an et~al. 2005). As part of the analysis (see below), we classify and
analyze sources in several control fields
to obtain an accurate estimate of the expected contamination for 
each galaxy in the survey
(Peng et~al. 2006ab; Jord\'an et~al. 2007, in preparation).

GC catalogs for each galaxy, combined
with similar information for the likely contaminants, will be used to study the 
color distribution of GCs (cf. Peng et~al. 2006a),
their distribution of sizes (cf Jord\'an et~al. 2005),
their luminosity functions (cf. Jord\'an et~al. 2006; 2007) and their
distribution within the color-magnitude diagram (cf. Mieske et~al. 2006b).
With the addition of  many new Fornax galaxies, it will be possible to take a
first look into the role played by environment in shaping the overall properties of
GC systems. The ACSFCS observations will also provide a direct test
of the accuracy of the median GC half-light radii as a standard ruler for  distance
estimation (Jord\'an et~al. 2005). We  note that the sizes of GCs
surrounding NGC~1399 derived from the ACSFCS data have already been used by Mieske et~al. (2006a) to provide additional evidence for the onset of a GC mass-size relation at 
masses of $M \gae 2 \times 10^6 M_\odot$, as originally suggested by 
Ha\c{s}egan et al. (2005).

It had been recognized since the early days of X-ray astronomy that GCs are
highly efficient in producing low-mass X-ray binaries (LMXBs; Clark 1975; Katz 1975). 
Thanks to the excellent imaging capabilities of Chandra and HST, tremendous progress has  been made in recent years on the connection between LMXBs and GCs in external galaxies as distant as 
$\sim 30$ Mpc. Such studies have shed light
on the processes that form LMXBs in dense GC environments and have  also
produced some surprising results, such as the fact that metal-rich GCs are $\approx 3$
times more efficient at forming LXMBs than their metal-poor counterparts (see, e.g., 
Fabbiano 2006 and references therein). By combining new and archival data from {\it Chandra} with optical
data from the ACSVCS we have carried out comprehensive studies of the connection between GCs and LMXBs in Virgo galaxies (Jord\'an et~al. 2004a; Sivakoff et~al. 2006). The  ACSFCS data will enable such studies to be extended to the Fornax cluster.

\begin{figure*}
\plotone{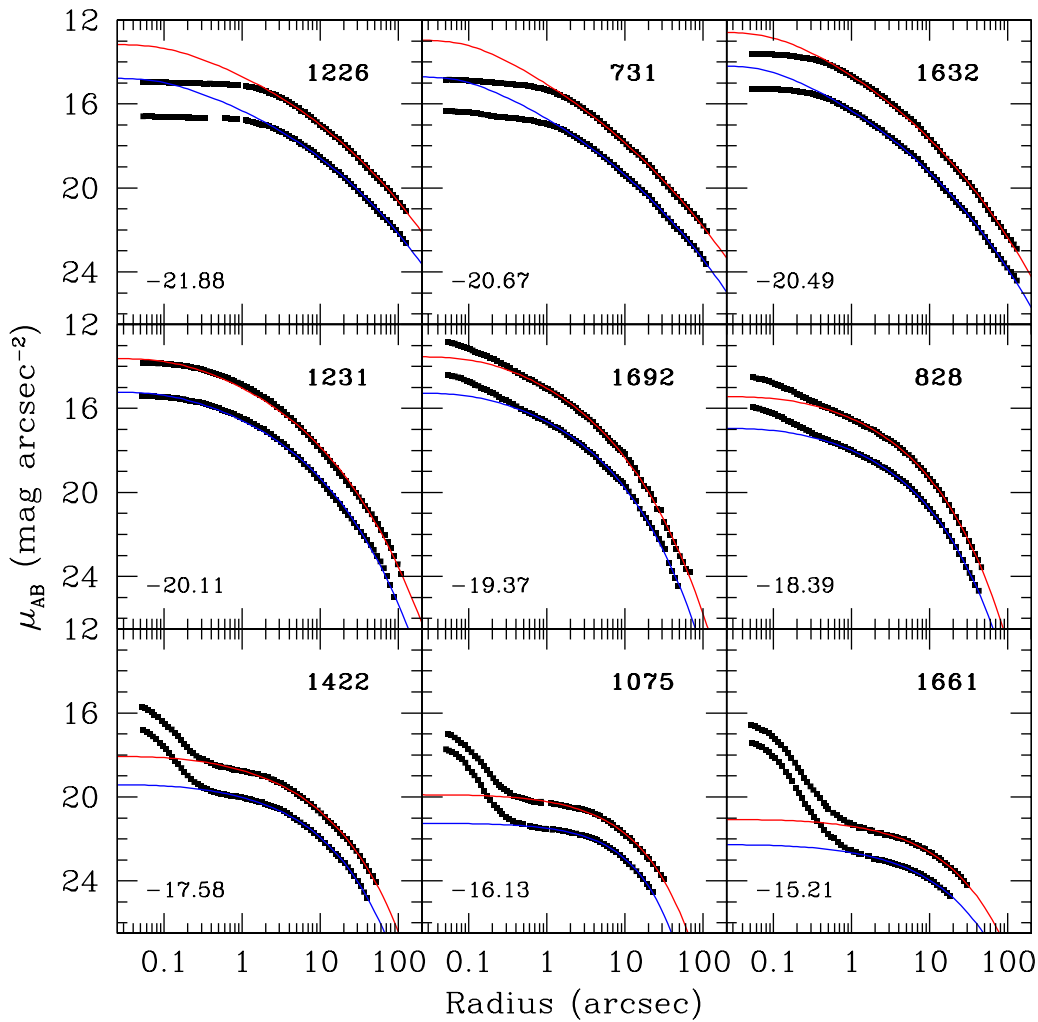}
\caption{Representative surface brightness profiles for nine early-type
galaxies from the ACSVCS spanning a factor of $\sim 460$ in blue luminosity; the $B-$band magnitude of each galaxy is listed in the corresponding panel.
In each panel, we show the azimuthally-averaged brightness profile in the
$\g$ and $\z$ bands plotted as a function of mean geometric radius (lower and upper profiles, respectively). The solid curves
show S\'{e}rsic models fitted to the profiles beyond $\sim 0\farcs2$-$2$\arcsec.
Note the gradual progression from a central light ``deficit" to ``excess", with
a transition at $M_B \sim -20$ (see Ferrarese et~al. 2006a and C\^ot\'e et~al.
2006 for details).
\label{fig:fcs_profiles}}
\end{figure*}

\subsection{Surface Brightness Profiles and the Core Structure of Early-Type Galaxies}

Surface photometry and isophotal studies have historically played an important role in 
shaping our understanding of the origin and structure of early-type galaxies (see, e.g., 
Kormendy \& Djorgovski 1989). In particular,
high-resolution imaging from HST has made it possible
to probe the innermost regions of nearby galaxies
allowing the systematic study of brightness profiles down to scales of
tens of parsecs. These inner profiles continue to hold much interest as it is now understood
that the physical processes at the centers of galaxies --- and, in particular, those
related to the supermassive black holes (SBHs) thought to reside there (Ferrarese \& Ford 2005) ---
are somehow connected to the history of the galaxy as a whole (e.g., Croton et~al. 2006).

The ACSFCS data will be used to measure surface brightness profiles and perform 
a study of dust morphology and nuclear properties for galaxies in a new and different 
environment (cf. Ferrarese et~al. 2006a, C\^ot\'e et~al. 2006). 
An interesting result that was unanticipated at the outset of the 
ACSVCS was the realization that previous ground-based studies of
early-type galaxies had significantly underestimated the number of galaxies which
contain compact stellar nuclei at, or near, their photocenters. It was found that
$\approx$ 70--80\% of the ACSVCS sample galaxies contained such nuclei,
roughly three times higher than previously believed (C\^ot\'e et~al. 2006).
This incidence of nucleation is similar to that found for late-type galaxies,
which often contain a ``nuclear star cluster" at, or near, their photocenters
(e.g., Carollo et al. 1997, 1998; Matthews et~al. 1999; B\"oker et al. 2002, 2004;
Walcher et al. 2005; Seth et~al. 2006).

The ACSVCS has allowed an exploration of the connection between 
core structure, stellar nuclei, and SBHs in new levels of detail.
On subarcsecond scales (i.e., $r \sim 0\farcs1$-$1\arcsec$, or $\sim$ 10-100 pc), 
the surface brightness profiles were found to vary systematically as one moves 
down the luminosity function (Ferrarese et al. 2006a; C\^ot\'e et~al. 2006). Bright
galaxies ($M_B \lesssim -20$), which in agreement with previous HST studies (e.g., Crane  et al. 1993; Ferrarese et~al. 1994; Lauer et~al. 1995) exhibit a nearly constant surface brightness cores, have surface brightness profiles that fall below the inward extrapolation of the S\'ersic model fitted 
beyond a few arcseconds (see also Graham et~al. 2003). Meanwhile, progessively fainter galaxies
show increasingly steep upturns over the S\'ersic models that fit the galaxies
on scales greater than $\sim$ $0\farcs1$-$1\arcsec$. In other words, on small
(subarcsecond) angular scales, galaxies were found to exhibit a gradual progession from a light ``deficit" to a light ``excess" (see Figure~\ref{fig:fcs_profiles}), while on larger scales (i.e., $\gtrsim$ a few arcseconds), the curved brightness profiles of real galaxies (both giants
and dwarfs) are accurately captured by S\'ersic models, but not by
the broken power-law parameterizations used in earlier studies (the ``Nuker law", Lauer et al. 1995).

Stellar nuclei are consistenly absent only in the brightest galaxies --- 
the same galaxies which are believed to host SBHs.  In addition, it was found that nuclei in the low- and intermediate-mass galaxies contribute a mean fraction, $\eta \sim 0.2\%$, of the total galaxy luminosity (C\^ot\'e et al. 2006,  Ferrarese et al. 2006b). This fraction is, to within 
the errors, the same as the fractional {\it mass} contribution of the central SBHs in 
massive early-type galaxies (see also Rossa et al. 2006 for similar conclusions
regarding nuclear star clusters in late-type galaxies).
Long-slit, integrated-light spectra for several dozen galaxies in the
ACSVCS were used to derive dynamical galaxy masses ${\cal M}_{\rm gal}$. Combining
these masses with masses for the nuclei derived from the brightness
profiles and stellar population models, it was found that a single 
${\cal M}_{\rm CMO}$-${\cal M}_{\rm gal}$~relation
extends smoothly from SBHs to nuclei as one moves down the mass
function for early-type galaxies (Ferrarese et al. 2006b; 
see also Wehner \& Harris 2006). This suggests that a single
mechanism, perhaps star formation or SBH accretion caused by gas inflows,
may be responsible for the growth and/or formation
of both types of objects (Ferrarese et~al. 2006b).
It also suggest that galaxy mass may be the primary
(though not necessarily only) parameter regulating such growth.
The ACSFCS data will be used to determine the extent to which galaxies in Fornax 
obey these new scaling relations, and to examine the ``universality" of the
relations defined by the sample of Virgo galaxies.

\subsection{New Families of Hot Stellar Systems}

Recent years have seen the discovery of stellar systems occupying regions of 
star cluster parameter space that had not been well explored before. 
One such class of objects are the so-called
ultra-compact dwarfs (UCDs; Hilker et~al. 1999; Drinkwater et~al. 2000), 
apparently old stellar systems
that are characterized by 
masses $M \gae 5 \times 10^6 M_{\odot}$ 
and sizes $r_e \lae 50$ pc (see Ha\c{s}egan et~al. 2005 
and references therein, where the term
Dwarf-Globular Transition Objects [DGTOs] was introduced to
avoid biasing the classification of these unusual objects as either
dwarf galaxies or star clusters). 
In Virgo we found that some of the DGTOs show rather large sizes
as compared to GCs,
$r_h \sim 20$ pc, and $V$-band mass-to-light ratios in the range
6--9 (in solar units), 
while some objects show properties which are consistent with those of
bright GCs (Ha\c{s}egan et~al. 2005). Thus, DGTOs seem to be  a 
mixture of GCs and bona-fide UCDs that show evidence 
for high mass-to-light ratios. 

Another type of stellar system, characterized by their diffuse nature
and thus termed Diffuse Star Clusters (DSCs; Peng et~al. 2006b), were
uncovered in significant numbers in the course of the ACSVCS.
They have $g$-band surface brightness $\mu_g \gae 20$ mag arcsec$^{-2}$. 
These star clusters include the ``faint fuzzy'' clusters found in  
nearby lenticular galaxies by Brodie \& Larsen (2002).
At least 12 galaxies
in Virgo  contained a significant population of DSCs; nine of these  galaxies 
are classified as S0s. 
The ACSFCS will make it possible to carry out systematic searches for DGTOs and DSCs in a second galaxy cluster, allowing a further characterization of their properties and perhaps providing
new clues as to the role of environment in their formation and  evolution.

\subsection{The 3-D Structure of Galaxy Clusters and an Improved Calibration of
the SBF Distance Method}

Accurate distances for individual galaxies greatly enhance the power of a 
survey such as the ACSVCS or ACSFCS not only because it is then possible to measure
quantities in absolute units rather than relative ones, but also because the study of the three 
dimensional structure of the cluster has strong implications for models of structure formation.
The method of surface 
brightness fluctuations 
(SBF; Tonry \& Schneider 1988) --- in which distances are derived from the
ratio of the second and first moments of their stellar luminosity functions ---
offers an efficient and accurate means of measuring such distances. As with
the ACSVCS, the SBF method will be used to derive distances for the
ACSFCS program galaxies and to use HST/ACS to provide 
a new and improved calibration of the SBF technique.

In Mei et~al. (2005ab) we have described the data reduction procedures
appropriate for SBF measurements with the ACS. Those papers
demonstrated the feasibility of these measurements ---
a non-trivial issue due to the strong geometric distortion
of the ACS which requires drizzling the data to an undistorted frame.
The interpolations necessary in the last step can affect the power-spectrum
of the fluctuations, something that will depend on the kernel used
for the drizzling. But as shown in Mei et~al. (2005a), the measurements
are feasible in spite of this complication.

However, in order to measure absolute distances with the SBF method, a calibration 
of the dependence of the fluctuation magnitude on the stellar population
content of the galaxy (typically parametrized as an integrated color) is needed. In Mei et~al. (2005b), we
presented a calibration of the SBF method for the $\g$ and $\z$ bandpasses 
chosen for the survey and used the resulting distances to examine
Virgo's three dimensional structure (Mei et ~al. 2007).

\begin{figure}
\plotone{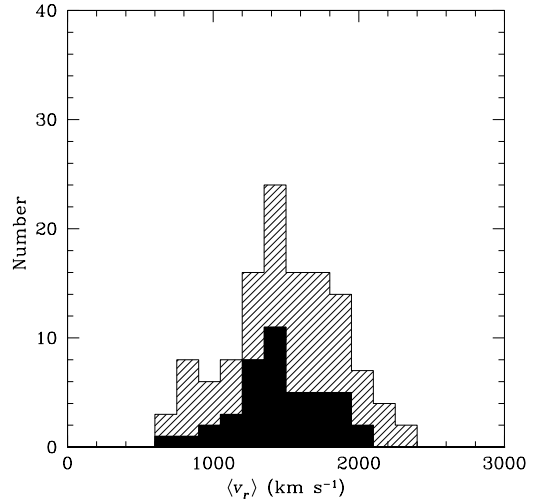}
\caption{Histogram of radial velocities for 125 galaxies
of all morphological types  identified as likely Fornax cluster
members by Ferguson (1989a) and having measured
radial velocities according to the NASA Extragalactic Database
({\it hatched histogram}). The filled histogram
shows the radial velocity distribution of the program
galaxies of the ACS Fornax Cluster Survey.
\label{fig:fcs_radvel}}
\end{figure}

The Fornax cluster certainly shows some substructure that will be interesting
to examine with new and better distances (e.g., Drinkwater et~al. 2001a; Dunn \& Jerjen 2006).
Relative to Virgo, however, the amount of substructure is quite modest. This feature of
Fornax, along with its more compact nature, makes it an obvious target for a 
program to calibrate the SBF relation (with the added benefit that well-calibrated 
$I$-band SBF distances exist for 26 Fornax galaxies; Tonry et~al. 2001).

\section{Sample Selection}
\label{sec:selec}

\begin{figure*}
\plotone{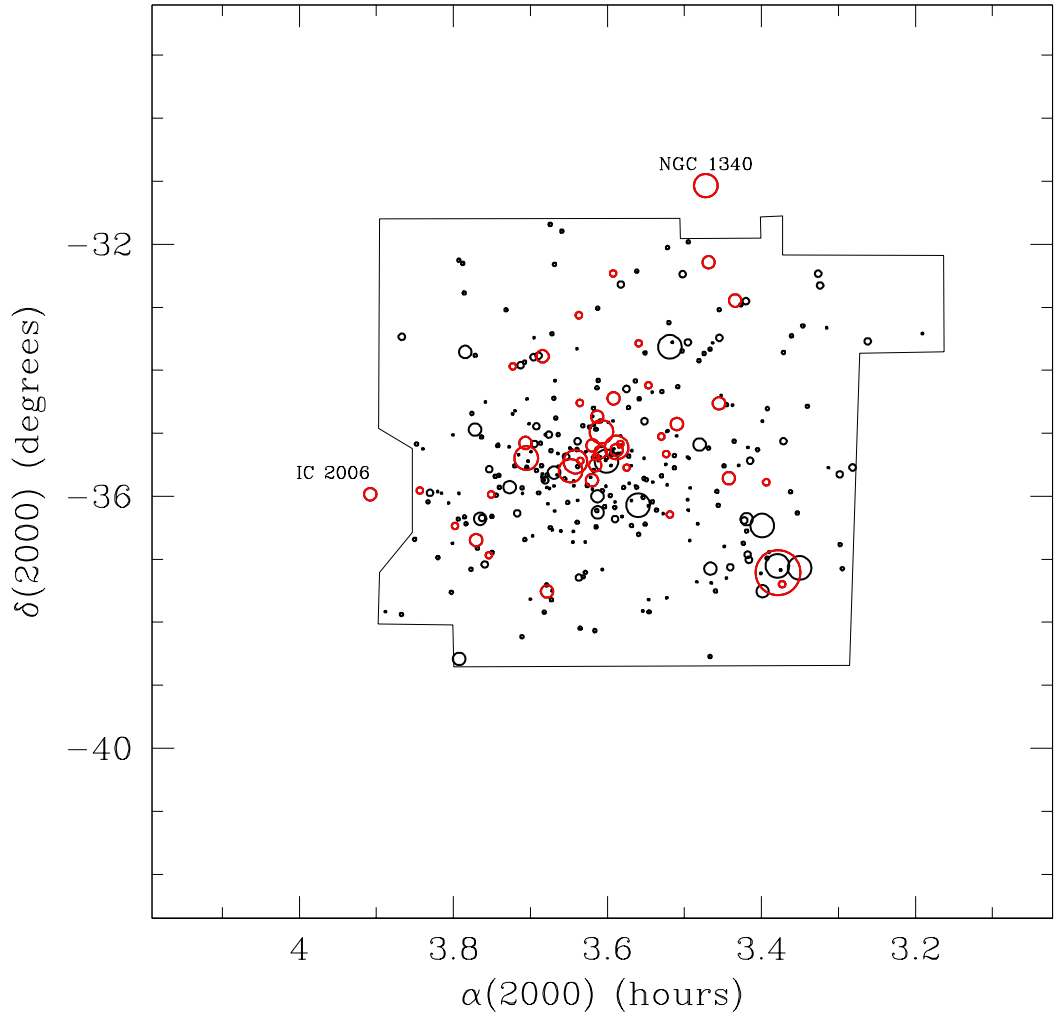}
\caption{Distribution of Fornax Cluster Catalog (FCC) galaxies on the plane of the sky. The 
outlined region is the area surveyed in constructing the FCC (Ferguson 1989a).
The symbol size is proportional to the galaxies' blue luminosity.
This figure shows the 340 galaxies within the FCC survey limits that are classified as likely members 
(i.e., membership codes 1 and 2) of the Fornax cluster, with no restriction on morphological type.
Red symbols  denote the full sample of 43 early-type galaxies
from the ACS Fornax Cluster Survey. Two galaxies that are part of the
ACS Fornax Cluster Survey but lie outside of the FCC survey region are 
labeled (NGC1340 and IC2006). 
The scale of this figure is the same as Figure~2 in C\^ot\'e et~al. (2004).
\label{fig:fcs_sky}}
\end{figure*}

Target galaxies were selected from the Fornax Cluster 
Catalog (FCC; Ferguson 1989a), which is based on wide-field blue plates
from the 2.5m du Pont telescope at Las Campanas Observatory and digitized plates
from the ESO/SRC survey of the southern sky. The Ferguson (1989a) survey, 
which remains the most complete and homogeneous
available for Fornax, covers an area of $\sim 40$ deg$^2$ centered at
$\alpha \sim 3^h35^m$ and $\delta \sim -35.7\degr$. Within the
survey area, the catalog contains 340 likely cluster members. Memberships
were established mainly through galaxy morphology, luminosity and surface
brightness, supplemented where possible by a small set of radial velocities.

Among the 340 likely members of Fornax, 79 galaxies
have $B_T \leq 15.5$, which is the adopted faint-end cutoff of the 
ACSFCS.  Early-type galaxies were selected
from this subset using the FCC morphological classifications.
Specifically, program galaxies were required to have
morphological types E, S0, SB0, dE, dE,N, dS0, or 
dS0,N\footnote{Including two S0/a transition type 
galaxies: \object[FCC 167]{FCC167} and \object[FCC 152]{FCC152}.}. 
This selection leaves a total of 44 galaxies. 
Additionally, we further checked that the galaxies were not classified
as late-types in NED; this step eliminated \object[FCC 338]{FCC~338} which is 
classified as type ``Sab: sp'' in NED. Because one of the main scientific
goals of the ACSFCS is an accurate calibration
of the $z$-band SBF distance indicator, two
Fornax galaxies that lie outside the FCC survey area were
added to the sample  (\object[NGC 1340]{NGC~1340}\footnote{NGC1340 also appears as NGC1344 in the
New General Catalogue (of Nebulae and Clusters).} and \object[IC 2006]{IC~2006}).
As a total of 44 orbits were awarded by the TAC, it was necessary
to drop one more galaxy in order to accommodate these two additions; as
a result, \object[FCC 135]{FCC~135} was excluded as it listed by NED as the fainter 
of the galaxies that have $B_T=15.5$ according to the FCC.

Unfortunately, the final planned 
observation of the survey --- that of \object[FCC 161]{FCC~161} = NGC~1379 ---
was not completed due to a failure of the telescope
to acquire the necessary guide stars and the shutter to open.
Thus, the final ACSFCS sample consists of an essentially complete sample of
41 early-type Fornax cluster galaxies brighter than $B_T \sim 15.5$
($M_B\sim -16$) mag, plus the outlying elliptical galaxies \object[NGC 1340]{NGC~1340}
and \object[IC 2006]{IC~2006}.

According to NED, radial velocities are now available for 125 of the 340
FCC galaxies classified as likely Fornax members by Ferguson (1989a).
The hatched histogram of Figure~\ref{fig:fcs_radvel}
shows the distribution of these 125 galaxies, while the filled
histogram shows the distribution of radial velocities for the sample 
of 43 ACSFCS galaxies. A Kolmogorov-Smirnov test shows that
both distributions are consistent with being drawn from the same
parent distribution ($p$-value $=0.91$) so we conclude that 
the velocity distribution for the ACSFCS program
galaxies is representative of the full sample of FCC galaxies with 
available radial velocities.

Figure~\ref{fig:fcs_sky} shows the distribution of the FCC 340 likely
Fornax members on the sky (black circles). For comparison,
the red circles show the 43 galaxies from the ACSFCS. The scale
of this Figure is the same as Figure~2 of C\^ot\'e et~al. (2004), which 
shows similar data for the Virgo cluster. A comparison of these
two figures clearly illustrates the more compact and regular nature
of Fornax.

\begin{figure}
\plotone{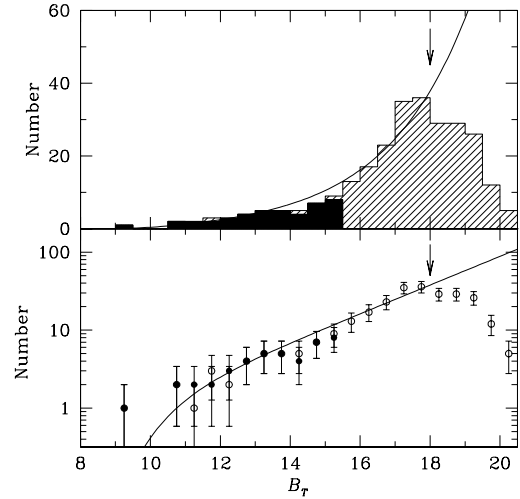}
\caption{({\it Upper panel}) Luminosity function of 269
early-type galaxies that are classified by Ferguson (1989a)
as members of the Fornax cluster ({\it upper, hatched histogram}).
The arrow shows the FCC completeness limit, while the 
solid curve shows the best-fit Schechter function
for E+S0+dE+dS0 galaxies from Ferguson \& Sandage (1991).
The filled lower histogram shows the luminosity function
for the 43 early-type galaxies in the ACS Fornax Cluster Survey.
({\it Lower panel}) Same as above, except in logarithmic form.
The open circles show the luminosity function of 
269 early-type members of the Fornax cluster according to
Ferguson (1989a). The arrow shows the completeness limit of the
FCC, while the filled circles show the luminosity function
of galaxies in the ACS Fornax Cluster Survey. The solid curve
is the same as that shown in the upper panel.
\label{fig:lf}}
\end{figure}

The upper panel of Figure~\ref{fig:lf} shows the luminosity function
of 269 early-type galaxies judged by Ferguson (1989a) to be members of Fornax; the lower panel
of this figure shows the same luminosity function in logarithmic 
form\footnote{For the luminosity function of 
early-types in Fornax at fainter galaxy magnitudes than those observed by 
Ferguson (1989a) see Mieske et~al. 2006c}. 
The filled
histogram ({\it upper panel}) and filled circles ({\it lower panel}) show the corresponding
luminosity functions for the ACSFCS program galaxies, which have
$9.4 \le B_T \le 15.5$, corresponding to a factor of $\approx 275$ in blue luminosity.
Note that these galaxies are all considerably brighter than the FCC completeness limit
of $B_T \approx 18$ (indicated by the arrows in both panels).

The general properties of the sample galaxies are presented in 
Table~\ref{tab:data}. From left to right, the columns of this table 
record the magnitude ranking from 1 to 43 (which also serves
as the identification number for each galaxy in the survey),
the FCC, NGC or IC identifier, right ascension and declination, 
total blue magnitude $B_T$ from Ferguson (1989a), radial velocity, 
morphological type from the FCC,
alternative names in the NGC
or ESO catalogs, and references for the radial velocity.\footnote{Magnitudes
and morphological types for \object[IC 2006]{IC~2006} and \object[NGC 1340]{NGC~1340} 
are taken from NED.}

\section{Observations}
\label{sec:obs}

Observations for each of the 43 program galaxies were carried out as part of GO program 10217 between September 2004 and March 2005. Each galaxy was imaged within a single orbit,
using the ACS 
Wide Field Channel (WFC) mode (Ford et~al. 1998; Sirianni et~al. 2005). 
This camera consists of two butted 2048$\times$4096 CCD detectors
(15$\mu$m pixels) having spectral response in the range 0.35--1.05$\mu$m.
The center of each galaxy was positioned near the
WFC aperture, at pixel position (2096,~200) on the 
WFC1 detector, and then offset 
perpendicular to the gap between the WFC1 and WFC2 detectors. 
For the 5  brightest
galaxies, this offset was 10$^{\prime\prime}$. 
An offset of 5$^{\prime\prime}$ was applied to the remaining galaxies. 
After correcting for geometrical distortion (see \S\ref{sec:data_red}), 
our images have a size of $4256 \times 4256$ pixels,  
with a pixel scale of $0\farcs049$.   
Thin wedges along the sides of the final
image and the $\approx 2\farcs5$ gap separating the WF1 and WF2 chips
are not illuminated, the wedges because of geometric distortion. 
The resulting field of view is approximately 
$202\arcsec \times 202\arcsec$ in the shape of a rhomboid.

\begin{deluxetable*}{lcllccllc}
%\rotate
\tabletypesize{\scriptsize}
\tablecaption{Basic Data for ACS Fornax Cluster Survey Galaxies.\label{tab:data}}
\tablewidth{0pt}
\tablehead{
\colhead{ID} & 
\colhead{Name} & 
\colhead{$\alpha$ (J2000)} &
\colhead{$\delta$ (J2000)} &
\colhead{$B_T$} &
\colhead{$v_r$} &
\colhead{Type} & 
\colhead{Other} & 
\colhead{ref} \\
\colhead{} &
\colhead{} &
\colhead{} &
\colhead{} &
\colhead{(mag)} &
\colhead{(km s$^{-1}$)} &
\colhead{}  &
\colhead{}  &
\colhead{} \\
\colhead{(1)} &
\colhead{(2)} &
\colhead{(3)} &
\colhead{(4)} &
\colhead{(5)} &
\colhead{(6)} &
\colhead{(7)}  &
\colhead{(8)}  &
\colhead{(9)} 
}
\startdata
1 & \object[FCC 21]{FCC~~21} &  03 22 42.09 & -37 12 31.63 & 9.4 & $1760\pm 10$ &   S0  (pec)                  &   NGC1316 & 1 \\
2 & \object[FCC 213]{FCC~213} &  03 38 29.14 & -35 27 02.30 & 10.6 & $1425\pm 4$ &   E0                         &   NGC1399 & 2 \\
3 & \object[FCC 219]{FCC~219} &  03 38 52.08 & -35 35 37.67 & 10.9 & $1947\pm 4$ &   E2                         &   NGC1404 & 2 \\
4 & \object[NGC 1340]{NGC~1340} &  03 28 19.7 & -31 04 05 & 11.3 & $1169\pm 15$ & E5 & NGC1344 & 3\\
5 & \object[FCC 167]{FCC~167} &  03 36 27.45 & -34 58 31.09 & 11.3 & $1877\pm 12$ &   S0/a                       &   NGC1380 & 4 \\
6 & \object[FCC 276]{FCC~276} &  03 42 19.16 & -35 23 36.03 & 11.8 & $1388\pm 3$ &   E4                         &   NGC1427 & 2 \\
7 & \object[FCC 147]{FCC~147} &  03 35 16.74 & -35 13 33.94 & 11.9 & $1294\pm 2$ &   E0                         &   NGC1374 & 2 \\
8 & \object[IC 2006]{IC~2006} &  03 54 28.45 & -35 58 01.7 & 12.2 & $1382\pm 3$ & E & ESO359-G07 & 5\\
9 & \object[FCC 83]{FCC~~83} &  03 30 35.04 & -34 51 14.51 & 12.3 & $1514\pm 3$ &   E5                         &   NGC1351 & 5 \\
10 & \object[FCC 184]{FCC~184} &  03 36 56.84 & -35 30 23.85 & 12.3 & $1302\pm 12$ &   SB0                        &   NGC1387 & 6 \\
11 & \object[FCC 63]{FCC~~63} &  03 28 06.55 & -32 17 05.87 & 12.7 & $1392\pm 3$ &   E4                         &   NGC1339 & 2 \\
12 & \object[FCC 193]{FCC~193} &  03 37 11.67 & -35 44 39.73 & 12.8 & $921\pm 12$ &   SB0  (5)                   &   NGC1389 & 7 \\
13 & \object[FCC 170]{FCC~170} &  03 36 31.59 & -35 17 43.35 & 13.0 & $1724\pm 9$ &   S0(9)(boxy)                &   NGC1381 & 4 \\
14 & \object[FCC 153]{FCC~153} &  03 35 30.91 & -34 26 44.74 & 13.0 & $1619\pm 6$ &   S0  (9)                    &   ESO358-G26 & 4 \\
15 & \object[FCC 177]{FCC~177} &  03 36 47.35 & -34 44 17.25 & 13.2 & $1561\pm 6$ &   S0  (9)(cross)             &   NGC1380A & 4 \\
16 & \object[FCC 47]{FCC~~47} &  03 26 31.97 & -35 42 44.59 & 13.3 & $1418\pm 3$ &   E4                         &   NGC1336 & 2 \\
17 & \object[FCC 43]{FCC~~43} &  03 26 02.30 & -32 53 36.80 & 13.5 & $1323\pm 17$ &   dS0 /2 (5),N               &   ESO358-G01 & 7 \\
18 & \object[FCC 190]{FCC~190} &  03 37 08.86 & -35 11 37.54 & 13.5 & $1740\pm 17$ &   SB0                        &   NGC1380B & 7 \\
19 & \object[FCC 310]{FCC~310} &  03 46 13.67 & -36 41 43.24 & 13.5 & $1373\pm 13$ &   SB0                        &   NGC1460 & 7 \\
20 & \object[FCC 249]{FCC~249} &  03 40 41.92 & -37 30 33.30 & 13.6 & $1613\pm 34$ &   E0                         &   NGC1419 & 8 \\
21 & \object[FCC 148]{FCC~148} &  03 35 16.79 & -35 15 55.95 & 13.6 & $740\pm 6$ &   S0(cross)                  &   NGC1375 & 4 \\
22 & \object[FCC 255]{FCC~255} &  03 41 03.40 & -33 46 38.42 & 13.7 & $1255\pm 23$ &   S0  (6),N                  &   ESO358-G50 & 3 \\
23 & \object[FCC 277]{FCC~277} &  03 42 22.60 & -35 09 10.22 & 13.8 & $1640\pm 8$ &   E5(boxy)                   &   NGC1428 & 2 \\
24 & \object[FCC 55]{FCC~~55} &  03 27 17.90 & -34 31 29.17 & 13.9 & $1279\pm 17$ &   S0(9),N                    &   ESO358-G06 & 7 \\
25 & \object[FCC 152]{FCC~152} &  03 35 33.09 & -32 27 44.79 & 14.1 & $1389\pm 12$ &   S0/a pec                   &   ESO358-G25 & 7 \\
26 & \object[FCC 301]{FCC~301} &  03 45 03.49 & -35 58 16.95 & 14.2 & $1007\pm 18$ &   E4                         &   ESO358-G59 & 3 \\
27 & \object[FCC 335]{FCC~335} &  03 50 36.64 & -35 54 29.27 & 14.2 & $1430\pm 2$ &   E                          &   ESO359-G02 & 2 \\
28 & \object[FCC 143]{FCC~143} &  03 34 59.06 & -35 10 09.90 & 14.3 & $1334\pm 2$ &   E3                         &   NGC1373 & 2 \\
29 & \object[FCC 95]{FCC~~95} &  03 31 24.68 & -35 19 46.41 & 14.6 & $1275\pm 26$ &   dSB0 or dSBa               &   & 7 \\
30 & \object[FCC 136]{FCC~136} &  03 34 29.39 & -35 32 41.18 & 14.8 & $1205\pm 1$ &   dE2,N                      &   & 2 \\
31 & \object[FCC 182]{FCC~182} &  03 36 54.24 & -35 22 22.69 & 14.9 & $1657\pm 19$ &   SB0 pec                    &   & 7 \\
32 & \object[FCC 204]{FCC~204} &  03 38 13.60 & -33 07 31.29 & 14.9 & $1369\pm 28$ &   dS0(8),N                   &   ESO358-G43 & 7 \\
33 & \object[FCC 119]{FCC~119} &  03 33 33.73 & -33 34 17.84 & 15.0 & $1374\pm 7$ &   S0   pec                   &   & 2 \\
34 & \object[FCC 90]{FCC~~90} &  03 31 08.06 & -36 17 19.48 & 15.0 & $1813\pm 15$ &   E4 pec                     &   & 7 \\
35 & \object[FCC 26]{FCC~~26} &  03 23 37.16 & -35 46 38.68 & 15.0 & $1823\pm 29$ &   SB0  (8)                   &   ESO357-G25 & 3 \\
36 & \object[FCC 106]{FCC~106} &  03 32 47.62 & -34 14 14.18 & 15.1 & $2064\pm 35$ &   d:S0(6),N                  &   & 7 \\
37 & \object[FCC 19]{FCC~~19} &  03 22 22.77 & -37 23 45.55 & 15.2 & $1497\pm 47$ &   dS0  (8),N                 &   ESO301-G08                       & 7 \\
38 & \object[FCC 202]{FCC~202} &  03 38 06.40 & -35 26 17.96 & 15.3 & $808\pm 22$ &   d:E6,N                     &   NGC1396 & 7 \\
39 & \object[FCC 324]{FCC~324} &  03 47 52.61 & -36 28 13.25 & 15.3 & $1856\pm 29$ &   dS0  (8)                   &   ESO358-G66 & 3 \\
40 & \object[FCC 288]{FCC~288} &  03 43 22.61 & -33 56 14.78 & 15.4 & $1189\pm 29$ &   dS0(9),N                   &   ESO358-G56 & 3 \\
41 & \object[FCC 303]{FCC~303} &  03 45 13.93 & -36 56 07.63 & 15.5 & $1980\pm 31$ &   dE1,N                      &   & 7 \\
42 & \object[FCC 203]{FCC~203} &  03 38 09.10 & -34 31 01.08 & 15.5 & $1138\pm 28$ &   dE6,N                      &   ESO358-G42 & 7 \\
43 & \object[FCC 100]{FCC~100} &  03 31 47.52 & -35 03 00.72 & 15.5 & $1660\pm 31$ &   dE4,N                      &   & 7 \\

\enddata
\tablecomments{
Key to columns---(1) Galaxy ID; 
(2) FCC, NGC or IC identifier; 
(3)-(4) Right ascension and declination; 
(5) Total $B_T$ magnitude from Ferguson (1989a);
(6) Radial velocity from the reference indicated in the last column and obtained via NED;
(7) Morphological classification from Ferguson (1989a);
(8) Alternative NGC or ESO identifier when available;
(9) Reference for radial velocity: 
1.- Longhetti et~al. (1998); 
2.- Graham et~al. (1998); 
3.- de Vaucouleurs et~al. (1995) (RC3); 
4.- D'Onofrio et~al. (1995);
5.- Smith et~al. (2000); 
6.- Menzies, Coulson \& Sargent (1989) ; 
7.- Drinkwater et~al. 2001b; 
8.- da Costa et~al. (1998).
}
\end{deluxetable*}

For each galaxy, five images were taken using an identical
observing sequence: $i.e.$, 
two 380 sec exposures in the F475W filter (760 sec total integration in F475W), two 565 sec
exposures in the F850LP filter, and a single 90 sec exposure in F850LP
(1220 sec total integration in F850LP). 
The choice of filters was dictated by the long baseline, which results
in good sensitivity to metallicity and age of stellar populations, 
and by the fact that the $z$-band has good characteristics for SBF measurements; 
the rationale for the choice of filters is thoroughly documented
in \S4.1 of C\^ot\'e et~al. (2004). 
In order to remove chip defects and bad pixels
a line dither with a spacing of $0\farcs 146$ was performed in the 
pair of identical exposures of each filter.
For some of the program galaxies, the central
surface brightness in the redder bandpass can approach
${\mu}_z$(AB) $\simeq 12$~mag~arcsec$^{-2}$, so the 90 sec F850LP exposure
was required to repair saturated inner regions  in the deeper  images.
The entire dataset for each galaxy therefore consists of an identical
set of images which were reduced and analyzed as 
described below (\S~\ref{sec:data_red}). Except for the slightly
larger exposure times and the choice of a line dither pattern, the 
observational setting is the same as that of the ACS Virgo Cluster Survey.

Table~\ref{tab:obslog} gives the observing log for all ACS observations related
to program GO-10217. From left to right, the columns of this table record
the identification number of each program galaxy, the Fornax Cluster Catalog
number from Ferguson (1989a) (or NGC and IC identifiers), the universal date
of the observation, the dataset name, the universal time at
the start of each observation and the position angle, $\Theta$, of the
y axis of the WFC1 detector. The final two columns give the exposure
time and filter for each observation. NGC and ESO identifiers for those
galaxies that have them may be found in Table~\ref{tab:data}.

%\LongTables
\begin{deluxetable*}{llllccll}
%\rotate
\tabletypesize{\scriptsize}
\tablecaption{Log of Observations for GO-10217.\label{tab:obslog}}
\tablewidth{0pt}
\tablehead{
\colhead{ID} & 
\colhead{Name} & 
\colhead{Date} & 
\colhead{Dataset} & 
\colhead{UT} &
\colhead{$\Theta$} &
\colhead{$T$} &
\colhead{Filter}  \\
\colhead{} &
\colhead{} &
\colhead{} &
\colhead{} &
\colhead{} &
\colhead{(deg)} &
\colhead{(sec)} &
\colhead{}  
}
\startdata
1 & FCC~~21 &  2005 Feb 16 &  j90x01i7q &  03:09:30 &    85.95 &  90 & F850LP \\
 & &   &  j90x01i8q &  03:13:18 &    85.95 &  565 & F850LP \\
 & &   &  j90x01iaq &  03:25:19 &    85.95 &  565 & F850LP \\
 & &   &  j90x01icq &  03:37:57 &    85.95 &  380 & F475W \\
 & &   &  j90x01ieq &  03:46:53 &    85.95 &  380 & F475W \\
2 & FCC~213 &  2004 Sep 11 &  j90x02e6q &  07:59:03 &   281.46 &  90 &  F850LP \\
 & &   &  j90x02e7q &  08:02:51 &   281.46 &  565 &  F850LP \\
 & &   &  j90x02e9q &  08:14:52 &   281.46 &  565 &  F850LP \\
 & &   &  j90x02ebq &  08:27:30 &   281.46 &  380 &  F475W \\
 & &   &  j90x02edq &  08:36:26 &   281.46 &  380 &  F475W \\
\enddata
\tablenotetext{1}{Table \ref{tab:obslog} is presented in its entirety in
the electronic version of this paper. A portion is shown here for guidance
regarding its form and content.}
\end{deluxetable*}

\section{Data Reduction Procedures}
\label{sec:data_red}

\subsection{Image Reductions}
\label{ssec:image_red}

The ACSFCS data reduction procedures are
nearly identical to the ones adopted for the ACSVCS. The
customized data reduction pipeline developed to reduce the ACSVCS
data has been 
documented in detail in Jord\'an et~al. (2004b; hereafter J04).
Wherever possible we retained the identical
procedures for the two surveys to ensure the maximum
level of homogeneity.
In what follows, we briefly summarize the main steps in the
reduction process and describe in detail only those aspects
of the ACSFCS analysis that differ from the ones outlined
in J04.

The data reduction pipeline consists of a series of steps which are summarized
graphically in Figure~1 of J04. In the first of these, 
the raw images are registered using
a source matching algorithm (see \S2.2 of J04) and the empirically measured shifts are
then used with {\it multidrizzle} (Koekemoer et~al. 2002) to create geometrically corrected, 
cosmic-ray cleaned images of dimension 4256 $\times$ 4256 pixel 
(see \S\S 2.2, 2.3 in J04). Measuring accurate shifts between the images
is important to obtain the highest quality combined images (White 2006).

In contrast to the ACSVCS, the analysis for the ACSFCS will rely exclusively on
images drizzled with a ``Lanczos3'' kernel, as this kernel gives a sharper PSF in
the drizzled image. Given the unexpected but frequent occurence of marginally-resolved stellar 
nuclei in the ACSVCS galaxies (C\^ot\'e et~al. 2006; Ferrarese et~al. 2006ab), this
improvement in resolution for the slightly more distant Fornax sample is deemed
more important than the superior ability of the ``Gaussian'' kernel to repair defective
pixels (which was used in the ACSVCS for the isophotal analysis).
The value of the ``bits'' parameter, which indicates which pixels are to be 
considered good based on their value in the data quality file and thus drizzled into the final 
image,  was determined
according to the date the data were taken, as there was a change in the convention
that defines the bits value: for observations taken before October 1,  2004 we adopted 
{\tt bits} $=14594$ and {\tt bits} $=14658$ for all subsequent observations.

Weight images were constructed in order to perform object detection and to aid 
in the determination of photometric and structural parameters of GCs.
Weight images were constructed as described in \S2.4 of J04
except that a different contribution from SBF
to the ``noise''
was adopted in order to account for the higher mean distance to the Fornax cluster
compared to Virgo. This SBF ``noise'' is added to avoid detection of spurious sources
corresponding to real fluctuations, in particular in the $z$-band. Following the notation
of J04, the contribution of SBF to the weight images $W^\prime_{ij}$ was given 
by $\kappa O_{ij}$ with $\kappa=0.073$ and $2.19$ for the $g$- and $z$-band respectively.
These values were derived by assuming that Fornax is a factor of 1.17 more distant
than the Virgo cluster (Tonry et~al. 2001).

Internal dust obscuration in some galaxies, usually in the central regions, 
significantly affects the surface brightness profiles. This poses problems for both
object detection and estimation of the local background over which sources are detected.
We used the method described in \S2.4 of J04, and discussed in more detail 
in Ferrarese et~al. (2006a), to mask pixels affected by dust when necessary.
Pixels found to be affected by dust are then given zero weight in the 
weight images. The galaxies that needed masking because of dust obscuration
are \object[FCC 21]{FCC~21}, \object[FCC 119]{FCC~119}, 
\object[FCC 152]{FCC~152}, \object[FCC 167]{FCC~167}, 
\object[FCC 184]{FCC~184}, \object[FCC 219]{FCC~219}, 
\object[FCC 335]{FCC~335} and \object[FCC 90]{FCC~90}.

A two-dimensional galaxy model for each program galaxy was then constructed in 
order to determine the weight maps and, more importantly, to subtract the galaxy
light and perform object detection on a nearly flat background. As in the ACSVCS (see \S2.5 in J04)
most galaxies were modeled using the ELLIPROF program described in the SBF
survey of Tonry et~al. (1997), masking regions affected by dust when necessary. 

However, for three of the galaxies, ELLIPROF was unable to produce acceptable models because
of the presence of strong edge-on disk components that were not well approximated
by elliptical isophotes modulated by low-order Fourier terms\footnote{These galaxies
are \object[FCC 153]{FCC~153}, \object[FCC 170]{FCC~170} and 
\object[FCC 177]{FCC~177}.}. In these cases, we masked the disk with a
rectangular region and then modeled the galaxy outside this region by using the
multi-gaussian expansion algorithm of Cappellari (2002) for 
\object[FCC 153]{FCC~153} and \object[FCC 177]{FCC~177}
and by using SExtractor to fit a two-dimensional bicubic spline for 
\object[FCC 170]{FCC~170}. The
masked regions are considered no further in the analysis of the star clusters.

\begin{figure}
\plotone{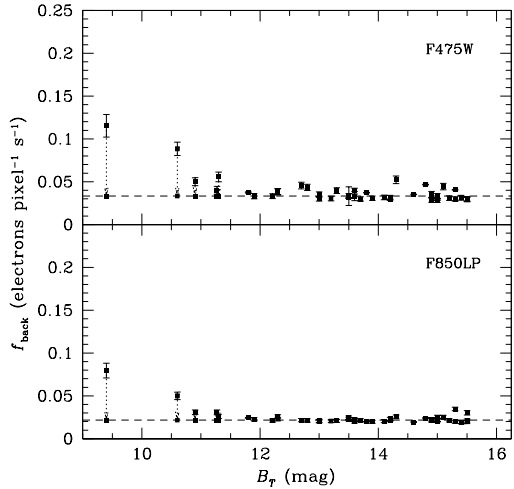}
\caption[]{Background count rates measured at $R \gae 90\arcsec$ in the F475W and F850LP images(upper and lower panels, respectively). 
Apart from the brightest galaxies, the background levels
are well represented by the values $f_{\rm back}=0.03324$ and $f_{\rm back}=0.02159$ 
electrons pixel$^{-1}$ s$^{-1}$ in the F475W and F850LP bands, respectively. These values 
are indicated by the dashed lines in each panel.
\label{fig:sky_1}}
\end{figure}

An important step in constructing the galaxy models is the determination of the
background ``sky'' rates, $f_{\rm back}$, for each galaxy.  Figure~\ref{fig:sky_1}
shows the measured count rates in 
F475W and F850LP, in units of electrons pixel$^{-1}$ s$^{-1}$,
for the full sample of galaxies. These
count rates are obtained by estimating the mode at distances 
$\approx 90\arcsec$--$120\arcsec$ from the galaxy center. The upturn
seen for a few of the brighter galaxies is due to the fact that the galaxy 
itself fills the field of view, thereby biasing the background estimation.

\begin{figure}
\plotone{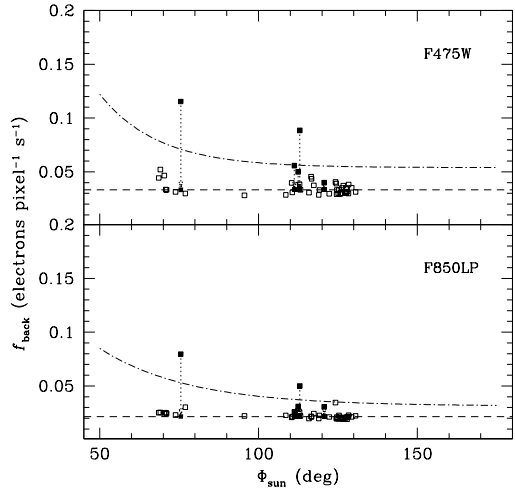}
\caption[]{Background count rates in F475W (upper panel) and F850LP (lower panel) plotted as a function
of $\Phi_{\rm Sun}$. The sky levels show no clear dependence on $\Phi_{\rm Sun}$. The dashed lines in 
the two panels show sky levels of $0.03324$ and $0.02159$ electrons pixel$^{-1}$ s$^{-1}$ in F475W and
F850LP, respectively. These values were used to estimate the background for the five brighter
galaxies in the sample. The dot-dashed curve show a parametric representation of the form of 
eq.~(5) in Jord\'an et~al. (2004) which was the adopted form for the background rates as a 
function of $\Phi_{\rm Sun}$ for the ACS Virgo Cluster Survey.  The measured count rates for the five
brightest galaxies are shown by the upper filled squares; the adopted values are indicated by the
lower filled squares.
\label{fig:sky_2}}
\end{figure}

In the ACSVCS, it was found 
that the local background scaled with the angle $\Phi_{\rm Sun}$ between
the Sun and the V1 axis of the telescope. That lead us to fit a function
of the form $f_{\rm back} = a_i\exp^{b_i(\Phi_{\rm Sun}-c_i)} + d_i$ to the 90 faintest
VCS galaxies and then use the best fit values of the parameters $a_i, b_i, c_i$ and $d_i$
to predict $f_{\rm back}$ for the bright galaxies with biased background estimates.
 In the case of the ACSFCS, there is no obvious dependence
of the background level on $\Phi_{\rm Sun}$, as illustrated in Figure~\ref{fig:sky_2}.
Therefore, to estimate the background for the five brighter
galaxies in the sample we adopted constant background levels of
$0.03324$ and $0.02159$ electrons pixel$^{-1}$ s$^{-1}$
for F475W and F850LP, respectively. These values are indicated
by the dashed lines in Figure~\ref{fig:sky_2}. The dot-dashed curves show
the models adopted in the ACSVCS, illustrating that the background levels are lower
for galaxies in Fornax\footnote{Of the four faint ACSFCS galaxies that show
higher backgrounds in F475W two (including the one showing the highest background)
have luminous companions within $\sim 5\arcmin$: \object[FCC 143]{FCC~143} is projected close to
\object[FCC 147]{FCC~147} and 
\object[FCC 202]{FCC~202} to \object[FCC 213]{FCC~213}=NGC~1399.}. The difference in the level and
behavior of the background levels is likely due to the low ecliptic latitude
of Virgo ($\beta\approx14.4^\circ$) as compared to that of Fornax ($\beta\approx-52.8^\circ$).

Objects were detected by running SExtractor (Bertin \& Arnouts 1996) 
on images in which the background (including the galaxy
light) was subtracted using the procedures detailed in \S2.5 of J04.
Sources were detected independently in the two bands and then matched
using a matching radius of $0\farcs1$. Beginning with the complete list of 
detections, a first selection on magnitude and elongation is then made
to isolate potential GC candidates for further analysis,
as described in \S2.6 of J04. The only difference in this procedure in
the case of the ACSFCS is that, when selecting in 
magnitude, we assume that the peak of the GC luminosity
function is 0.34 mag fainter due to the larger relative distance of 
Fornax (i.e., $d_{\rm fornax}/d_{\rm virgo} = 1.17$; Tonry et~al. 2001).

All sources flagged as potential GCs are then run through a code (KINGPHOT) that
fits a PSF-convolved King (1966) model to the observed light distribution of each 
object (Jord\'an et~al. 2005).
This provides measurements of both the King structural parameters for the sources
as well as their total magnitudes. Performing such measurements
requires accurate models for the PSFs, something that proved 
challenging in the ACSFCS. These challenges, and the adopted solutions,
are described in detail below (\S\ref{ssec:psf}).

The pixel coordinates of the detections were first converted to celestial
coordinates using the header information. We then compared the
coordinates of 371 astrometric standards within our survey fields
to those listed in the Guide Star Catalog 2.3 (McLean et~al. 1998).
For 35 galaxies which had enough objects to derive a reliable correction,
we applied offsets in right ascension and declination to the
celestial coordinates derived from the headers. The mean corrections
were $-0\farcs016$ in right ascension and $0\farcs34$ in declination.
The internal accuracy of the celestial coordinates for a given galaxy
is $\approx 0\farcs01$ (Meurer et~al. 2002). 

A reddening for each galaxy was computed using the DIRBE maps of
Schlegel, Finkbeiner \& Davis (1998). The mean reddening for the whole sample was found
to be $\langle E(B-V) \rangle = 0.013$ mag 
with a dispersion around this value of $\sigma_{E(B-V)} = 0.003$~mag.
The extinction ratios and photometric zeropoints adopted are those
described in \S2.7 of J04 (which were derived using information
from Sirianni et~al. 2005). 

A consequence of the environment of HST on silicon-based CCDs 
is to decrease their charge transfer efficiency (CTE) with time, 
especially along the serial direction (e.g., Riess \& Mack 2004). 
CTE degradation  has the effect of reducing
the apparent flux of objects, which in turn potentially requires corrections to the
photometry in order to recover the ``true" flux of objects. We have used
the CTE corrections for ACS/WFC presented in Riess \& Mack (2004) to estimate the effect of CTE 
degradation on our survey. The worst case scenario is offered by a source at the 
detection limit in $\g$ (26.35), observed at the latest date our data were taken and subject to 
2048 serial pixel transfers. In this case we expect charge transfer inefficiency to dim the source 
by  $\approx 0.025$ mag. 
A typical source with $\g =24$, observed at the average date for our survey and
subject to 1024 serial pixel transfers would be dimmed by $\approx 0.004$ mag.
Given the modest magnitude of the flux loss in the worst case scenario and the 
small effect of CTE degradation for a typical source we choose not to apply CTE correction
to our photometry.

The final step in generating GC catalogs is the selection of candidate GCs from the
set of objects/detections that passed the initial selection criteria on magnitude and
elongation. This selection is done using a mixture model algorithm that
separates probable GCs from contaminants, assigning to each source a probability
$p_{\rm GC}$ that it is indeed a GC. This selection procedure
is described in detail in Jord\'an et~al. (2007, in preparation). This
paper also includes a discussion of the methods by which the aperture
corrections are determined for the GCs.

\subsection{Point Spread Function and Globular Cluster Size Measurements}
\label{ssec:psf}

\begin{figure}
\plotone{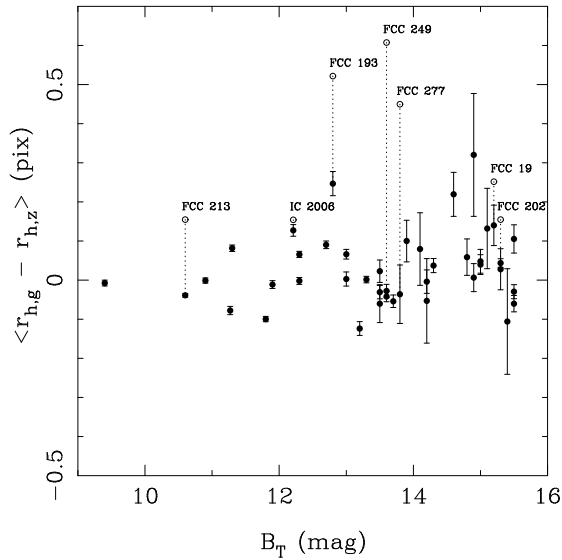}
\caption[]{Median difference $\langle r_{h,g} - r_{h,z} \rangle$ between the half-light radii $r_h$ measured in the $g$- and $z$-bands as
 a function of total galaxy $B_T$ magnitude (solid points). The
mean difference is estimated as described in \S\ref{ssec:psf}. Open symbols denote the galaxies
that were flagged as having an anomalously large difference in the mean size and whose GC
sizes were re-measured using a corrected PSF built using on-frame stars (see \S\ref{ssec:psf}).
The FCC identification numbers of those galaxies is indicated next to the open symbols. The
dashed lines connect to the measured $r_h$ difference after the correction was performed.
\label{fig:rh_diff}}
\end{figure}

After an initial run of KINGPHOT on the detected GC candidates, it was noticed that,
in some galaxies, the mean half-light radii $\langle r_{h,g} \rangle $ in the
$g$-band was significantly larger than the corresponding value $\langle r_{h,z} \rangle $
in the $z$-band. In the ACSVCS, differences
in $\langle r_{h} \rangle $ between the two bands satisfied 
$|\langle r_{h,g} - r_{h,z}\rangle | \la 0.1$ pixel (see Jord\'an et~al. 2005). 
However, for some ACSFCS galaxies 
this difference was significantly larger: up
to $\approx 0.5$ pixels in a few cases (see Figure~\ref{fig:rh_diff}). As the underlying objects
are the same, the sizes in the two bandpasses should be roughly equal in the average.

Given that a study of the GC size distribution function and a comparison with the 
Virgo results presented in Jord\'an et~al. (2005) is one of the main goals
of the ACSFCS, it was necessary to determine the root cause of this problem and/or to provide
a means for correcting it. Aside from the obvious challenge that this problem
might impose on the interpretation of the size distribution function, it could
lead to unacceptably large (and systematic) differences
in the GC magnitudes derived from the fitted models.

As a first attempt to solve the problem, a new set of PSFs was constructed using
$\gae 1000$ stars in fields located in the outskirts of the Galactic GC 47 Tucanae (programs
GO-10048 and GO-10375). We determined three different PSFs using observations taken
in September 2004, December 2004
and February 2005. By doing so, we generated an empirical PSF for each galaxy that
was determined no more
than two months from the ACSFCS observations. This procedure was motivated by the fact
that, on 20 December 2004, HST underwent a secondary mirror adjustment (which the 
observatory is regularly subjected to in order to correct systematic changes in the
relative position of the primary and secondary mirrors). 
In this particular case, the secondary mirror was moved by $4.16 \mu$m, a sizable displacement which 
%http://www.stsci.edu/hst/observatory/focus/mirrormoves.html
certainly changed the focus and affected the PSF (Krist 2003). Although the empirically
derived PSFs clearly reveal the post-movement PSF to show a more compact core,
a redetermination of the GC sizes showed that these new PSFs did
not solve the anomalous sizes found in the $g$-band. Nevertheless, the new PSFs
do constitute an improvement over the single PSF approach adopted for the ACSVCS (both
in terms of time sampling and number of stars used to construct the PSF) so
we kept them for the analysis of all galaxies that do not show an anomalous  size difference.
For galaxies that were observed after 20 December 2004, we used the PSF determined on February
2005; for the ones observed before, we used the closest in time of the other two PSFs.

While the cause of the problem was under investigation, Anderson \& King (2006)
published a comprehensive study of the WFC PSF, including an analysis of its time variability. They find
that the WFC PSF varies on orbital timescales in an unpredictable way, with variations 
in the core flux of up to $\sim 10\%$. Moreover, these variations are more
pronounced in the bluer filters as compared to a red one (see their Figure~8).
This closely resembles the behavior that is seen in the ACSFCS data --- core flux
variations of that order
in the $g$-band would easily result in the derived $r_h$ being systematically
higher by tenths of a pixel while being insignificant in the near-infrared
$z$-band. The rapid time variability of the effect is clearly illustrated by the
pair of giant elliptical galaxies 
\object[FCC 213]{FCC~213} and \object[FCC 219]{FCC~219}. Both galaxies contain sizable GC populations 
and thus have excellent statistics on the average size difference between the two bands. Yet,
despite the fact that the observations
were carried out within a day of each other, \object[FCC 213]{FCC~213} 
showed a difference of $\sim 0.15$ pixel while
that for \object[FCC 219]{FCC~219} was essentially zero. 

The results of Anderson \& King (2006) suggest that the only viable way to correct the wider
PSF in the $g$-band is to identify on-frame stellar objects and use them
to empirically correct the PSFs that were constructed from the 47 Tucanae fields.
Of course, the drawback of this procedure
is that the accuracy of any such correction will be limited by the number of suitable stars
that can be found. We proceeded by first identifying all galaxies that
required a correction. To do this, the measured $r_h$ in both bands
were used to compute $\langle r_{h,g} - r_{h,z}\rangle$ for a sample of objects satisfying
$\z < 23.5$ mag and $0\farcs01 < \langle r_h \rangle < 0\farcs05$ --- criteria that ensure
the sample is dominated by GCs (Jord\'an et~al. 2005). We then flagged those
galaxies for which 
$$\langle r_{h,g} - r_{h,z}\rangle > \max (0.1 \,{\rm pix}, 4\sigma) \eqno{(1)}$$
where $\sigma$ is the standard deviation of the mean. A total of seven galaxies were
identified in this way: 
\object[FCC 213]{FCC~213}, \object[IC 2006]{IC~2006}, 
\object[FCC 193]{FCC~193}, \object[FCC 249]{FCC~249}, 
\object[FCC 277]{FCC~277}, \object[FCC 19]{FCC~19} and \object[FCC 202]{FCC~202}.

For these galaxies, we adopted the following approach to
correct the PSFs. First, in each case, we identified from the SExtractor catalog
of the $z$-band image a set of probable stars by selecting objects with CLASS\_STAR $>0.8$, 
full-width at half-maximum FHWM $<2.5$ pixel and $z$-magnitude in a 4-pixel aperture
$z_{\rm ap} < 20$ mag. These objects were then inspected visually to 
verify the automated selection; this step resulted in the rejection of a single
star candidate in just one galaxy (FCC~19).
The number of stars found in each galaxy ranged from 5 to 8. Once these
stars were selected, we determined for each of the affected galaxies a perturbation to the 
$g$-band PSF to adjust it to the selected on-frame stars. That is to say, if we denote
the normal survey PSF by $\mathcal{P}_{ij}$, and the PSF that best describes the on-frame
stars (in a least squares sense) by $\mathcal{S}_{ij}$, then 
we determined for each of the $g$-band observations a perturbation $\delta_{ij}$
such that $\mathcal{S}_{ij} = \delta_{ij} + \mathcal{P}_{ij}$. Here $\delta_{ij}$ is 
an array of the average difference between the identified stars and the $\mathcal{P}_{ij}$.
For the affected galaxies,
sizes in the $g$-band were then determined using the perturbation $\delta_{ij}$ on top 
of the normal survey PSF. Given that the PSF quality is limited by the small number of stars
used to determine each perturbation, the $g$-band sizes in these galaxies are of somewhat
lower quality
than the corresponding $g$-band measurements in the remaining 36 galaxies.

The results of the procedure are summarized in Figure~\ref{fig:rh_diff} which
plots $\langle r_{h,g} - r_{h,z}\rangle$ 
against galaxy magnitude $B_T$ for the 42  galaxies for which we 
could measure this quantity\footnote{The galaxy \object[FCC 119]{FCC~119} was not left with enough
GC candidates after the cuts described above to calculate $\langle r_{h,g} - r_{h,z}\rangle$.}.
Open symbols indicate the initial measurements for those galaxies which were found to 
satisfy equation~(1). The dotted lines connect to solid points that
indicate the value of $\langle r_{h,g} - r_{h,z}\rangle$ using the corrected PSF.
As seen in the figure,
our procedure brings the behavior of $\langle r_{h,g} - r_{h,z}\rangle$ to acceptable levels
for most galaxies in the sample. We emphasize that at no point in the process do we 
{\it force} $\langle r_{h,g} - r_{h,z}\rangle$ to decrease. The improved agreement results
naturally from the perturbation to the PSF calculated using the on-frame stars selected from 
the $z$-band observations. After the correction procedure, we proceed with the analysis
knowing that the size measurements are consistent, to within $\la 0.1$ pixel,
between the two bands for the bulk of the ACSFCS sample.

\section{Summary}
\label{sec:summary}

We have provided a brief introduction to the ACS Fornax Cluster Survey
(GO-10217), an HST program to image 43 early-type galaxies brighter than
$M_B \approx -16$ in the Fornax cluster.
In conjunction with a similar survey of 100 early-type galaxies in the Virgo cluster --- 
the ACS Virgo Cluster Survey (C\^ot\'e et~al. 2004) --- the final sample of
F475W and F850LP imaging for 143 galaxies represents
the most comprehensive imaging dataset currently available for early-type galaxies in nearby cluster
environments. We have briefly described the observing strategy and the data reduction procedures
adopted for the survey. 
Scientific results from the survey will be presented in future papers
in this series.
Additional information on the ACS Fornax Cluster
Survey can be found at the program website:
\texttt{http://www.eso.org/~ajordan/ACSFCS/}.

\acknowledgements

Support for program GO-10217 was provided through a grant from the Space
Telescope Science Institute, which is operated by the Association of 
Universities for Research in Astronomy, Inc., under NASA contract NAS5-26555. 
P.C. acknowledges additional support provided by NASA LTSA grant NAG5-11714.
L.I. acknowledges additional support from FONDAP Center of Astrophysics.
This research has made use of the NASA/IPAC Extragalactic Database (NED)
which is operated by the Jet Propulsion Laboratory, California Institute
of Technology, under contract with the National Aeronautics and Space 
Administration. 

{\it Facility:} \facility{HST (ACS/WFC)}

\end{document}